\documentclass[a4paper,10pt]{article}
\usepackage[utf8]{inputenc}
\usepackage{graphicx}
\usepackage{txfonts}
\title{Studying the interstellar medium of H\,{\sc ii}/BCD galaxies using IFU spectroscopy}
\author{\it{Patricio Lagos and Polychronis Papaderos}\\
                                                    \\
       Centro de Astrof\'{\i}sica da Universidade do Porto \\ 
       Rua das Estrelas, 4150-762 Porto, Portugal\\
       plagos@astro.up.pt; papaderos@astro.up.pt
       }

\begin{document}

\maketitle

\begin{abstract}

We review the results from our studies, and previous published work, on the 
spatially resolved physical properties of a sample of H\,{\sc ii}/BCD galaxies, as obtained mainly from 
integral-field unit spectroscopy with Gemini/GMOS and VLT/VIMOS. 
We confirm that, within observational uncertainties, our sample galaxies show nearly spatially 
constant chemical abundances, similar to other low-mass starburst galaxies.
They also show He \,{\sc ii} $\lambda$4686 emission with properties being suggestive of a mix of excitation sources, 
with Wolf-Rayet stars being excluded as the primary one. 
Finally, in this contribution we include a list of all  H\,{\sc ii}/BCD galaxies studied thus far with 
integral-field unit spectroscopy.

\end{abstract}

\section{Introduction}

\begin{figure}
\begin{center}
\includegraphics[scale=0.4]{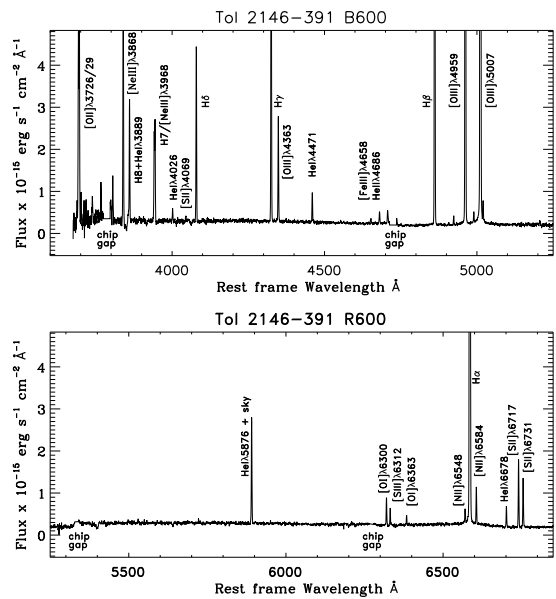} 
\caption{Integrated spectra of the H\,{\sc ii}/BCD galaxy Tol 2146-391. 
Top panel: blue spectrum (grating B600). Bottom panel: red spectrum (grating R600). 
}
   \label{fig1}
\end{center}
\end{figure}

The concept of \textit{compact galaxies} was introduced by Zwicky \cite{Zwicky64}, who has described them as
``galaxies barely distinguishable from stars" on the Palomar Sky Survey plates.  
The term blue compact dwarf (BCD) galaxies \cite{ThuanMartin81} identify those objects that show low luminosity, 
small linear dimensions, strong emission lines superposed on a blue continuum, and spectral properties 
that indicate low chemical abundances. 
BCDs form a subset of H\,{\sc ii} galaxies, a large number of which have been identified on objective prism surveys
by \cite{Haro56}, \cite{Zwicky66}, \cite{Markarian67} and \cite{Sargent70} by the presence of strong emission lines, 
similar to Giant H\,{\sc ii} regions (GH\,{\sc ii}Rs) in our galaxy. 
Here, we will refer to H\,{\sc ii}/BCD galaxies as objects with a metallicity 7.0$\leq$12+log(O/H)$\leq$8.4  
(e.g., \cite{KunthSargentS83}), low luminosity (M$_B$ $\gtrsim$ -18) and gas-rich objects (e.g., \cite{Filho13}) 
undergoing vigorous starburst activity in a relatively small physical size ($\lesssim$1 Kpc).  
The star-forming component, in these objects, typically contains multiple knots of star-formation with 
unresolved ensembles of young star clusters (e.g., \cite{Lagos07}, \cite{Lagos11}). 
The hypothesis of these systems being young, forming their first generation of stars  
has been discarded by the detection of an evolved underlying
stellar host with an age $>$1 Gyr, in the majority of the nearby H\,{\sc ii}/BCD population 
(e.g., \cite{Papaderos96},\cite{Telles97}). Figure \ref{fig1} shows the optical spectrum of the galaxy
Tol\ 2146-391 obtained using integral field unit (IFU) observations with Gemini/GMOS. 
In this figure we label the most important emission lines used in our studies, in particular, 
the strong Balmer hydrogen recombination lines and collisionally excited emission lines, 
such as [O \,{\sc ii}] $\lambda\lambda$3726,3729, 
[O \,{\sc iii}] $\lambda$4363, [O \,{\sc iii}] $\lambda$5007, [S \,{\sc ii}] $\lambda\lambda$6717,6731, 
[N \,{\sc ii}] $\lambda$6584, which have been used for the determination of physical conditions 
(e.g., electron temperature and density) and chemical abundances (e.g., oxygen, nitrogen, etc). 
We also detect in some of our galaxies the high-ionization emission line He \,{\sc ii} $\lambda$4686.

Although progress has been made in this field, important unsolved questions remain with regard to 
the mode of star formation (e.g., quasi-continuous vs fluctuating), and the triggering mechanism
of ongoing starburst activity in H\,{\sc ii}/BCDs.
It has been suggested \cite{Lagos11} that the cluster formation efficiency is lower in compact 
H\,{\sc ii}/BCD galaxies than the one found in more luminous galaxies.
These luminous systems generally show an irregular outer shape  
and kinematical signatures of merging in their interstellar medium (ISM). 
In some cases, the formation of star cluster complexes occurs coevally \cite{Lagos11},
whereas in others star formation occurs in a propagating manner \cite{Papaderos98,Papaderos08}.  
In any case, the mechanism which may trigger the current star-formation in these galaxies is not well understood;
in particular the relative importance of intrinsic and environmental properties remains a subject of investigation.

Another important issue is the chemical and kinematical imprints of star cluster formation and evolution 
on the spatially resolved properties of the ISM in H\,{\sc ii}/BCD galaxies.
As a natural consequence of star formation driven feedback, the newly synthesized elements
will be dispersed and mixed across the ISM via hydrodynamic mechanisms 
(e.g., \cite{Tenorio-Tagle96}), leading to \emph{chemical homogeneity in the oxygen abundance} \cite{Kobulnicky97,KobulnickySkillman98}
of H\,{\sc ii}/BCD galaxies (e.g., \cite{Lagos09} and references therein). 
The nitrogen-to-oxygen ratio N/O has also been found to be rather constant 
(log(N/O)$\simeq$-1.6; \cite{EdmundsPagel78,Alloin79,IzotovThuan99})  
at low metallicity (12+log(O/H)$\leqslant$7.6), suggesting primary production by massive stars \cite{IzotovThuan99}
as the main contribution to nitrogen enrichment at those very low metallicities.
A small fraction of H\,{\sc ii}/BCD galaxies fall in this very-low metallicity regime \cite{KunthOstlin00} 
and are commonly referred to as extremely metal poor (XMP) galaxies or XMP BCDs. 
These galaxies are the best nearby candidates for cosmologically young objects, as various arguments 
imply that they have formed most of their stellar mass in the past 1--3 Gyr \cite{Papaderos02}.
At intermediate metallicity (7.6$\leq$12+log(O/H)$\leq$8.3) the large observed spread in N/O 
has been attributed mainly to the loss of heavy elements via galactic winds \cite{Vanzee98}, 
and/or to the delayed release of nitrogen by intermediate and/or massive stars 
and oxygen by massive stars \cite{EdmundsPagel78, Garnett90}.
However, the delayed-release scenario cannot explain the presence of some H\,{\sc ii}/BCD 
galaxies with a high N/O ratio at low metallicities.
The most plausible explanation for the high N/O ratio observed in these
objects is the chemical pollution of the ISM by nitrogen released by 
massive Wolf-Rayet (WR) stars as is, apparently, the case of the well-studied BCD 
NGC 5253 \cite{WalshRoy89,Kobulnickyetal97,LopezSanchez07,Westmoquette13}.  
Finally, at higher metallicities (12+log(O/H)$\geq$8.3) the N/O ratio clearly increases 
with increasing oxygen abundance and the nitrogen content is mainly due to secondary production
by intermediate-mass stars.

\begin{figure}
\begin{center}
\includegraphics[scale=0.6]{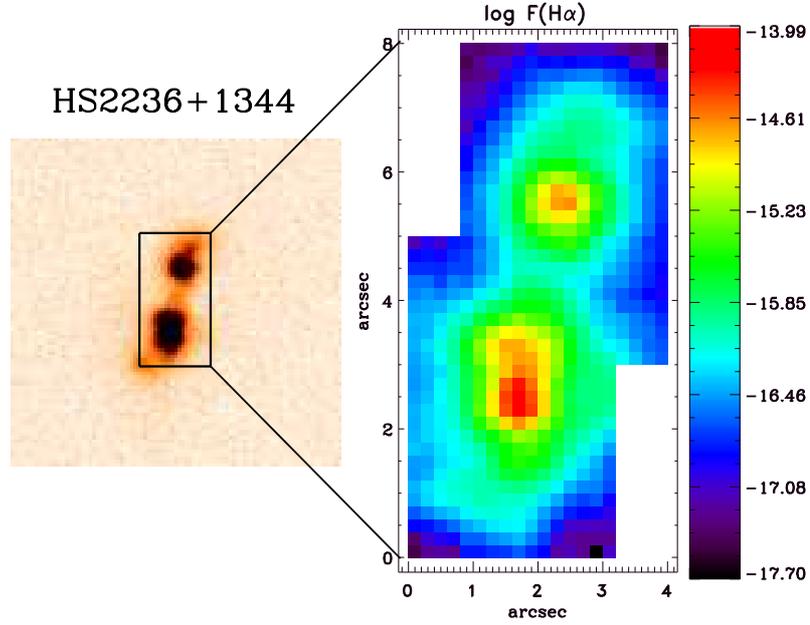} 
\caption{GMOS-IFU observation of the XMP galaxy HS2236+1344. Left: g-band acquisition image of the galaxy. 
The rectangle indicates the total field of view of 4"$\times$8" obtained from the composition of two different 
pointings with GMOS. 
Right: H$\alpha$ map of the galaxy. log F(H$\alpha$) in units of ergs cm$^{-2}$ s$^{-1}$.
Further details will be presented in a forthcoming paper \cite{Lagos13}.}
   \label{fig2}
\end{center}
\end{figure}

\begin{figure}
\begin{center}
\includegraphics[scale=0.6]{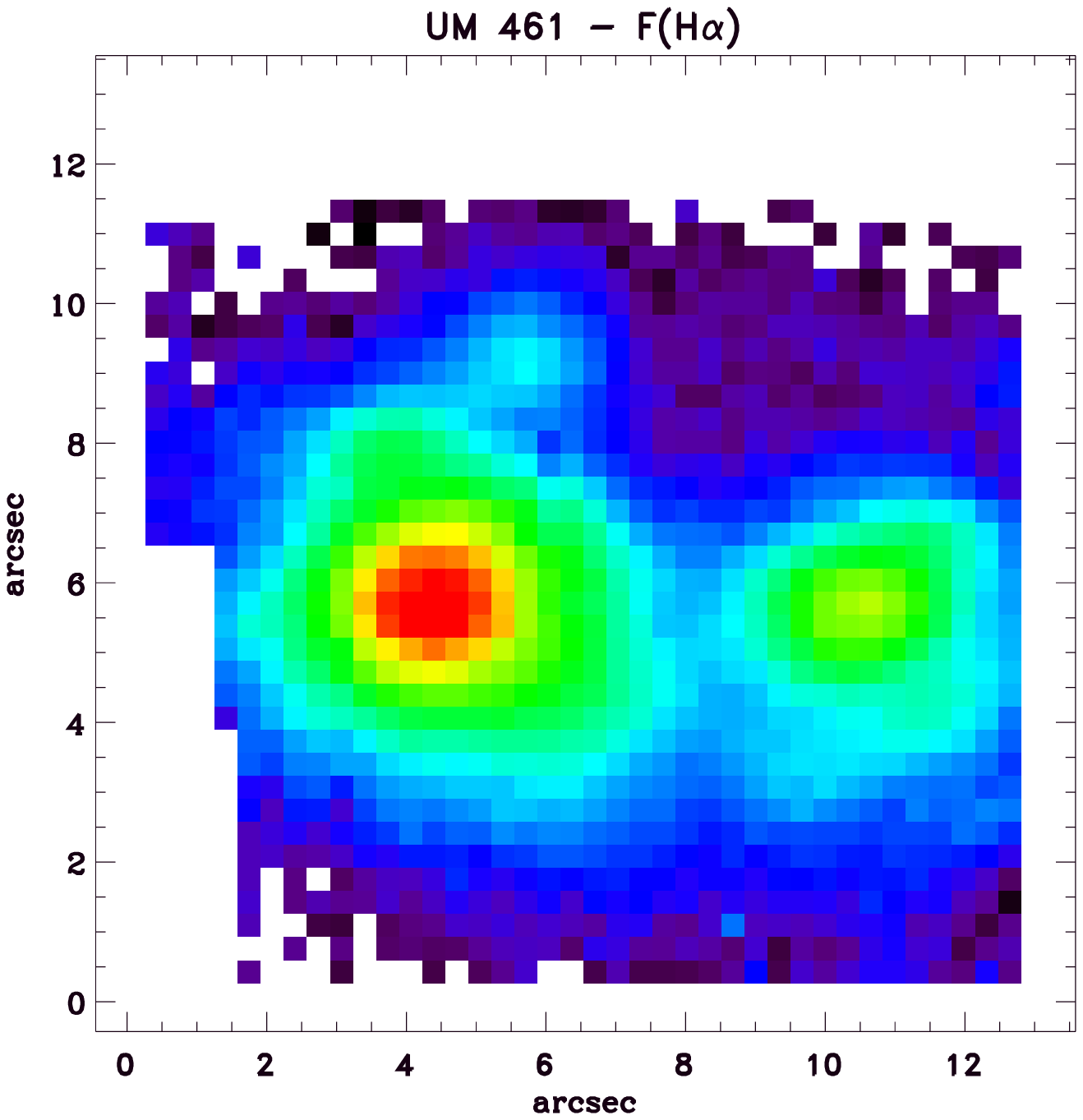} \\
\includegraphics[scale=0.6]{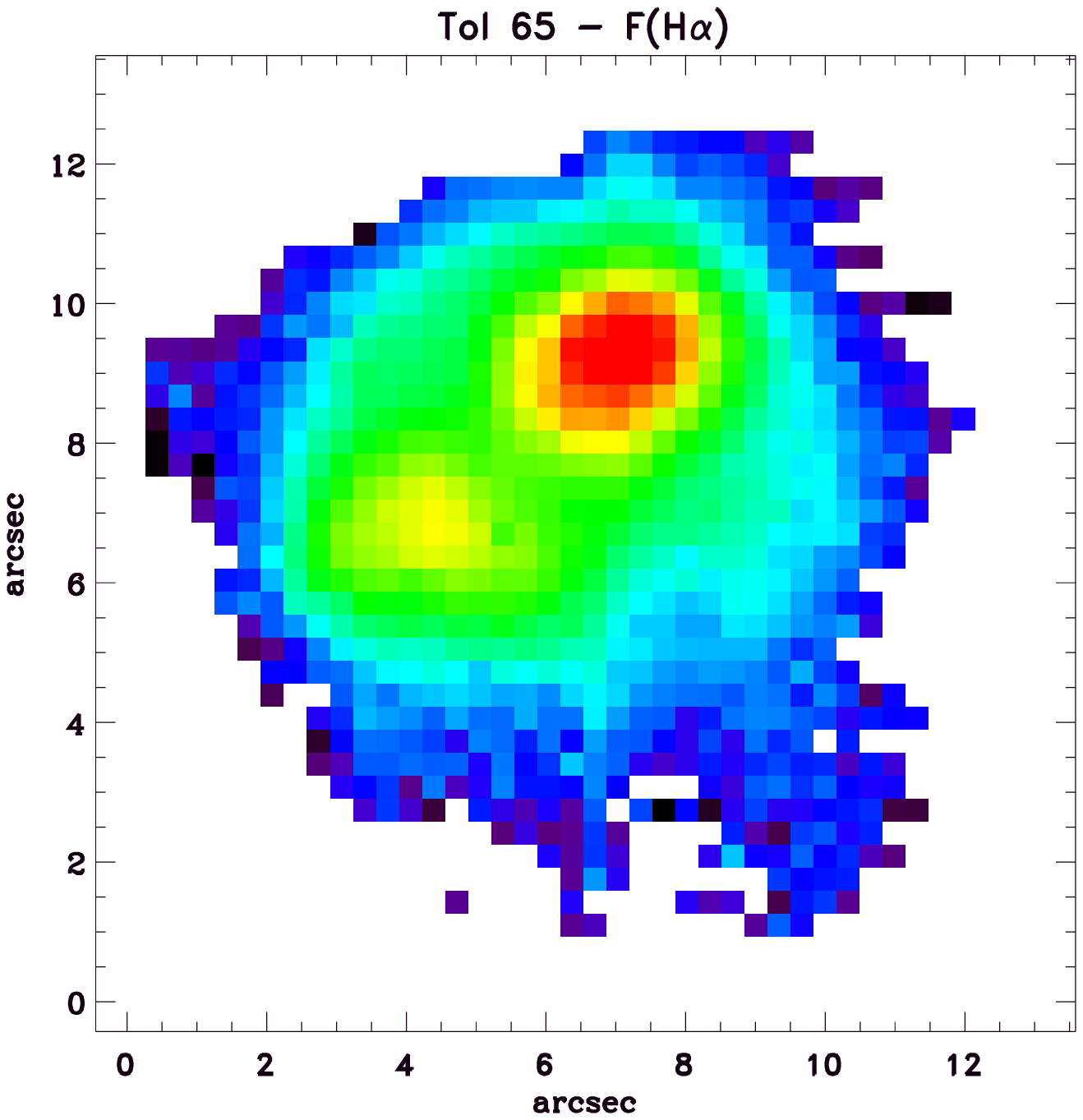} 
\caption{VIMOS-IFU H$\alpha$ emission line map of the XMP BCD galaxies UM\ 461 (Top panel) and Tol\ 65 (Bottom panel).
North is to the top and East to the left. Further details will be presented in a forthcoming paper.
}
   \label{fig3}
\end{center}
\end{figure}

So far, an increasing number of H\,{\sc ii}/BCD galaxies has been studied with IFU spectroscopy
(see Table \ref{tab1} where we provide an overview of the literature) with main focus on 
the spatial properties of the ISM.
Recently, we have started a program investigating with IFU spectroscopy the physical conditions 
in the ISM of the most compact H\,{\sc ii}/BCD galaxies, laying special emphasis on the 
extinction patterns, emission line ratios, oxygen and nitrogen
abundances, kinematics and the relation on the intrinsic properties of star formation as well as possible
evolutionary effects \cite{Lagos09,Lagos12}.
To this end, we observed a sample of H\,{\sc ii}/BCD galaxies using the GMOS-IFU on Gemini South and North 
and, more recently, with VLT/VIMOS.
The GMOS-IFU observations were performed using the gratings B600 and R600
in one slit mode, covering a total spectral range 
from $\sim$3000 to $\sim$7230 $\rm \AA$. This observational setup provides a pattern
of 500 hexagonal elements with a projected diameter of 0".2, covering a total
3".5 $\times$5" field of view (FoV).  
The VIMOS-IFU observations were obtained using the gratings HR$\_$blue and HR$\_$orange covering a 
spectral range from $\sim$3710 to $\sim$7700 $\rm \AA$. Our data yield a scale on the sky of 0."33 per fiber, 
and cover a FoV of 13" $\times$ 13". In Figure \ref{fig2} we show the g-band acquisition image of the
XMP BCD galaxy HS2236+1344, in which we indicated the total field of view of 4"$\times$8" and the H$\alpha$ map 
of the galaxy obtained from the composition of two different pointings with GMOS-IFU. 
In Figure \ref{fig3} we show the H$\alpha$ emission line map of the galaxies UM 461 and Tol 65 obtained using VIMOS-IFU.
Table~2 lists the general parameters of our sample of galaxies.

\begin{table*}
 \centering
 \small
 \begin{minipage}{140mm}
  \caption{H\,{\sc ii}/BCD galaxies with published IFU observations. The list could be incomplete 
 and does not include Fabry-Perot observations. Redshift distances obtained from NED assuming H$_{0}$ = 73 km s$^{-1}$ Mpc$^{-1}$.
 M$_B$ computed from the tabulated values of $m_{B}$ obtained from HyperLeda.
 }
  \begin{tabular}{@{}lccccccrlr@{}}
  \hline
   Name     &    \multicolumn{2}{c}{Coordinates (J2000)} &  Distance & M$_{B}$ & Reference & Instrument \\       
            &    \multicolumn{2}{c}{ }                   &   (Mpc)   &  (mag)  &           &         \\
   
 \hline
IC 10        & 00:20:17.3 & +59:18:14 &  0.8&-12.73   &\cite{LopezSanchez11}     & PMAS\\ 
Haro 11      & 00:36:52.7 & -33:33:17 & 84.6&-20.30   &\cite{James13b}           & FLAMES \\
Tol 0104-388 & 01:07:02.2 & -38:31:51 & 91.6&$\cdots$ &\cite{Lagos11b,Lagos12}   & GMOS-IFU\\ 
Mrk 996      & 01:27:35.5 & -06:19:36 & 22.2&-16.56   &\cite{James09}            & VIMOS-IFU\\
HS0128+2832  & 01:31:21.3 & +28:48:12 & 66.3&$\cdots$ &\cite{PerezMontero11}     & PMAS\\
UM 408       & 02:11:23.4 & +02:20:30 & 49.3&-15.76   &\cite{Lagos09,Lagos10}    & GMOS-IFU\\
UM 420       & 02:20:54.5 & +00:33:24 &240.1&-21.19   &\cite{James10}            &VIMOS-IFU\\
Mrk 370      & 02:40:29.0 & +19:17:50 & 10.8&-16.40   &\cite{GarciaLorenzo08}    & INTEGRAL\\
SBS 0335-052 & 03:37:44.0 & -05:02:40 & 55.6&-16.77   &\cite{Vanzi11,Izotov06a}  & SINFONI, FLAMES\\
IIZw 33      & 05:10:48.1 & -02:40:54 & 38.8&-18.45   &\cite{Cairos12}           & VIRUS-P\\
IIZw 40      & 05:55:42.6 & +03:23:32 & 10.8&-16.69   &\cite{Vanzi08,Bordalo09}  & SINFONI, GMOS-IFU\\
Haro 1       & 07:36:56.7 & +35:14:31 & 51.9&-21.02   &\cite{Cairos12}           & VIRUS-P\\
He 2-10      & 08:36:15.1 & -26:24:34 & 12.0&-17.93   &\cite{Marquart07,Cresci10}& FLAMES, SINFONI\\
HS0837+4717  & 08:40:29.9 & +47:07:10 &172.4&-18.15   &\cite{PerezMontero11}     & PMAS\\
Mrk 1418     & 09:40:27.0 & +48:20:15 & 10.6&-16.27   &\cite{Cairos09b}          & PMAS\\
Mrk 407      & 09:47:47.6 & +39:05:03 & 21.8&-16.50   &\cite{Cairos10}           & PMAS\\
Mrk 409      & 09:49:41.2 & +32:13:16 & 21.2&-17.11   &\cite{Cairos09a}          & PMAS\\
Tol 1004-296 & 10:06:33.4 & -29:56:05 & 15.3&-17.56   &\cite{Vanzi11}            & SINFONI\\
Mrk 32       & 10:27:02.0 & +56:16:14 & 11.4&-14.20   &\cite{Cairos10}           & PMAS\\
Mrk 35       & 10:45:22.4 & +55:57:37 & 12.9&-17.33   &\cite{GarciaLorenzo08}    & INTEGRAL\\
Mrk 178      & 11:33:28.9 & +49:14:14 &  3.4&-13.25   &\cite{Kehrig13}           & INTEGRAL\\
UM 448       & 11:42:12.4 & +00:20:03 & 76.3&-19.93   &\cite{James13a}           & FLAMES\\
Mrk 750      & 11:50:02.7 & +15:01:23 & 10.3&-14.30   &\cite{Cairos10}           & PMAS\\
UM 462       & 11:52:37.2 & -02:28:10 & 14.5&-16.20   &\cite{James10}            & VIMOS-IFU\\
Mrk 206      & 12:24:17.0 & +67:26:24 & 18.0&-15.90   &\cite{Cairos10}           & PMAS\\
NGC 4670     & 12:45:17.1 & +27:07:31 & 14.7&-17.68   &\cite{Cairos12}           &VIRUS-P\\
NGC 5253     & 13:39:55.9 & -31:38:24 &  5.6&-17.68   &\cite{Cresci10,MonrealIbero10,MonrealIbero12,Westmoquette13} & SINFONI, FLAMES, GMOS-IFU \\
Tol 1434+032 & 14:37:08.9 & +03:02:50 & 23.4&-14.94   &\cite{Cairos10}           & PMAS\\
Mrk 475      & 14:39:05.4 & +36:48:22 &  8.0&-13.10   &\cite{Cairos10}           & PMAS\\
IIZw 70      & 14:50:56.5 & +35:34:18 & 16.2&-16.28   &\cite{Kehrig08}           & PMAS \\
I Zw 123     & 15:37:04.2 & +55:15:48 &  9.1&-14.35   &\cite{Cairos10}           & PMAS\\
Mrk 297      & 16:05:13.0 & +20:32:32 & 65.0&-20.60   &\cite{GarciaLorenzo08}    & INTEGRAL\\
I Zw 159     & 16:35 21.0 & +52:12:52 & 37.0&-17.19   &\cite{Cairos10}           & PMAS\\
Tol 2146-391 & 21:49:48.2 & -38:54:09 &120.7& $\cdots$&\cite{Lagos11b,Lagos12}   & GMOS-IFU\\
HS2236+1344  & 22:38:31.1 & +14:00:30 & 84.5& $\cdots$&\cite{Lagos13}            & GMOS-IFU\\ 
Mrk 314      & 23:02:59.2 & +16:36:19 & 28.5&-18.20   &\cite{GarciaLorenzo08,Cairos12}& INTEGRAL, VIRUS-P\\
III Zw 102   & 23:20:30.1 & +17:13:32 & 22.4&-18.81   &\cite{GarciaLorenzo08,Cairos12}& INTEGRAL, VIRUS-P\\
Mrk 930      & 23:31:58.3 & +28:56:50 & 75.2&$\cdots$ &\cite{PerezMontero11}          & PMAS\\
\hline
\end{tabular}
\end{minipage}
\label{tab1}
\end{table*}

This contribution is organized as follows: the metallicity content and the spatial distribution 
of the ISM in our sample galaxies are described in Section \ref{metal}, and in Section \ref{hard} we discuss
the high-ionization emission line He \,{\sc ii} $\lambda$4686 and its relationship to the properties
of the ISM. Finally, in Section \ref{conclusion}, we summarize our results and conclusions.

\begin{table*}
 \centering
 \begin{minipage}{140mm}
  \caption{General parameters of our studied galaxies.}
  \begin{tabular}{@{}lccccccrlr@{}}
  \hline
   Name     &    \multicolumn{2}{c}{Coordinates$^{a}$ (J2000)}& D (Mpc)\footnote{Obtained from NED} & M$_{B}$\footnote{Obtained from HyperLeda} (mag)& 12+log(O/H) & Instrument \\       
   
 \hline

Tol 0104-388 & 01:07:02.2 & -38:31:51 &  91.6 & $\cdots$ &8.02$^{\cite{Lagos12}}$ & GMOS-IFU\\ 
UM 408       & 02:11:23.4 & +02:20:30 &  49.3 & -15.76   &7.87$^{\cite{Lagos09}}$ & GMOS-IFU\\
Mrk 600      & 02:51:04.6 & +04:27:14 &  13.8 & -15.38   &7.88$^{\cite{Guseva11}}$& VIMOS-IFU\\
UM 461       & 11:51:33.3 & -02:22:22 &  14.2 & -14.36   &7.72$^{\cite{Nava06}}$  & VIMOS-IFU\\
Tol 65       & 12:25:46.9 & -36:14:01 &  38.5 & -15.44   &7.53-7.56$^{\cite{Guseva11}}$ & VIMOS-IFU\\
Tol 2146-391 & 21:49:48.2 & -38:54:09 & 120.7 & $\cdots$ &7.82$^{\cite{Lagos12}}$ & GMOS-IFU\\
HS2236+1344  & 22:38:31.1 & +14:00:30 &  84.5 & $\cdots$ &7.55$^{\cite{Lagos13}}$ & GMOS-IFU\\ 

\hline
\end{tabular}
\end{minipage}
\label{tab2}
\end{table*}

\section{The metal content in the ISM of H\,{\sc ii}/BCD galaxies}\label{metal}

\begin{figure}
\centering
\includegraphics[scale=0.52]{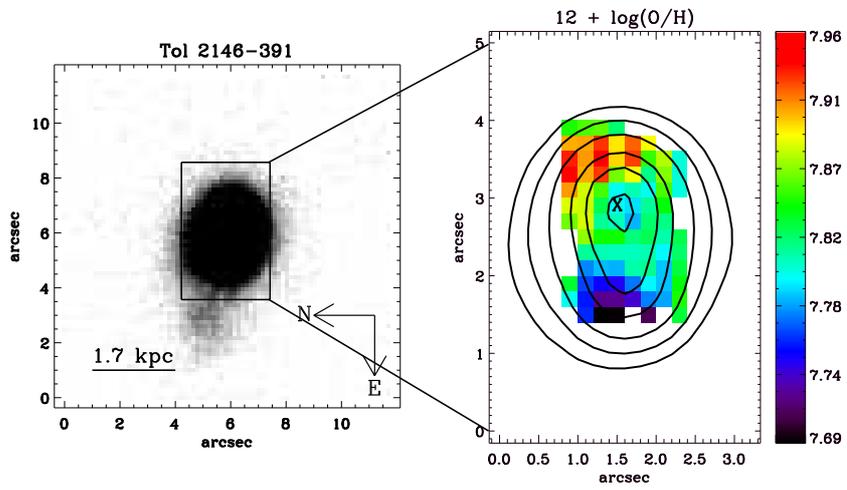} 
\caption{Spatial distribution of oxygen abundances in the galaxy Tol 2146-391. 
Left: g-band acquisition image of the galaxy. The rectangle indicates the FoV
of 3".5$\times$5" covered by our GMOS-IFU observation. Right: 12 + log(O/H) spatial distribution. 
The isocontours display the H$\alpha$ emission. 
The maximum H$\alpha$ emission is indicated in the maps by an X symbol.
We only considered spaxels with signal to noise ratio (S/N) $>$ 3 in the [OIII] $\lambda$4363 line.
More details in \cite{Lagos12}.}
   \label{fig4}
\end{figure}

Using the reddening corrected emission line intensities of the spectra of each one of the spaxels we can derive the
physical conditions (electron temperature and density) and the chemical abundances (O and N) across the ISM
of the galaxies. We calculate oxygen abundances in regions where the [OIII]$\lambda$4363 emission line has been 
detected assuming O/H = O$^{+}$/H$^{+}$ + O$^{++}$/H$^{+}$, while nitrogen abundances are obtained assuming
N/H = ICF(N) $\times$ N$^{+}$/H$^{+}$, with ICF(N) denoting the ionization correction factor (O$^{+}$+O$^{+2}$)/O$^{+}$.
For the sake of illustration, in Figure \ref{fig4} we show the spatial distribution of 12+log(O/H) 
in the GMOS-IFU FoV of the galaxy Tol 2146-391.
We can see in this figure that, despite a slight depression in the inner part of the galaxy,
the oxygen abundance appears to be uniform across the galaxy (see Figure 18 in \cite{Lagos12}).
In \cite{Lagos10} we compare the spatial distribution of 12+log(O/H), found in \cite{Lagos09}, with the position 
of the star cluster/complexes detected in the galaxy UM 408 by \cite{Lagos11} using high resolution near-IR  K$_{p}$-band images. 
We found that the variation of the observed data points (see  Figure 9 in \cite{Lagos09} and Figure 1 in \cite{Lagos10}) 
may not be statistically significant, indicating that these regions have identical chemical properties within the errors.
It is interesting to note that we observed a marginal gradient of decreasing abundance from the center outward in UM 408, 
indicating that the highest abundance values are found near the peak of H$\alpha$ emission and extinction c(H$\beta$), 
and coincident with the position of the brightest star cluster/complex \cite{Lagos11}. 
In any case, the absence of chemical overabundances in the ISM of UM 408, Tol 2146-391, Tol 0104-388 and HS2236+1344 
and in the dwarf galaxies studied in the literature 
(e.g., \cite{Skillman89,Kobulnicky97,KobulnickySkillman98,Lee06,Croxall09,Berg12}) 
indicates that the population of young star clusters 
is not producing localized oxygen overabundances. The most likely explanation for this is that 
metals formed in the current star-formation episode reside in a hot gas phase 
(T$\sim$10$^{7}$ K; \cite{Tenorio-Tagle96}); thus, they
are not observable in optical wavelengths. Whereas metals from previous star-formation events 
are well mixed and homogeneously distributed through the whole galaxy. 
In Tol 2146-391, the 12+log(N/H) radial distribution shows a slight decrease with radius. 
This would argue in favor of heavy elements being produced in a previous burst of star-formation and dispersed 
within the ISM by starburst-driven super-shells \cite{Lagos12}, while the depressed central region could be attributed 
to radial inflow of relatively low metallicity gas from large radii to the center, 
thus diluting the abundance of the gas in the nuclear region.

Regarding the integrated properties of the galaxies, \cite{Izotov06b}, \cite{Brinchmann08} and \cite{LopezSanchez10}
suggest that there is a dependence between N/O and the EW(H$\beta$), in the sense of an increasing N/O ratio 
with decreasing EW(H$\beta$). Izotov et al. \cite{Izotov06b} argue that this trend is naturally explained by 
nitrogen ejection from WR stars. 
In the following analysis, we mainly concentrate on the spatially resolved physical properties of the ISM  
in individual galaxies and their possible relation to the star formation process (e.g., the star formation history, 
burst parameter, WR star content).
In Figure 5 (see \cite{Lagos12}) we show the log (N/O) versus EW(H$\beta$) and 12+log(O/H) versus log (N/O) 
for all spaxels of the galaxies Tol 0104-388 and Tol 2146-391.
From that figure, it can be seen that the EW(H$\beta$) values are rather constant,
with a very small variation of equivalent widths as the N/O ratio increases. 
A comparison of log (N/O) versuss 12+log(O/H) in Tol 2146-391 (Figure 3b) reveals that
the log (N/O) values increase with the 12+log(O/H). 
This data point distribution has similar patterns to those found in H\,{\sc ii}/BCD galaxies by  \cite{IzotovThuan99}  
of increasing N/O ratios with respect to the oxygen abundance. 
The inner region of Tol 2146-391 (near the peak of H$\alpha$) presents N/O
ratios which are larger than those expected by pure primary production of nitrogen. 
This might be a signature of time delay between the release of oxygen and nitrogen \cite{KobulnickySkillman98}, 
or gas infall or outflow.
In any case, for the metallicity of Tol 2146-391 purely secondary nitrogen enrichment appears
implausible. 
In the case of HS2236+1433,  we reported in \cite{Lagos13} evidence for a high N/O ratio 
in one of the three GH\,{\sc ii}Rs of the galaxy.
But again, the spatial distribution of these abundances, at large scales, lead us to consider that oxygen and hydrogen
are well mixed and homogeneously distributed over the ISM of the galaxy.  

\begin{figure}
\begin{center}
\includegraphics[width=2.8in]{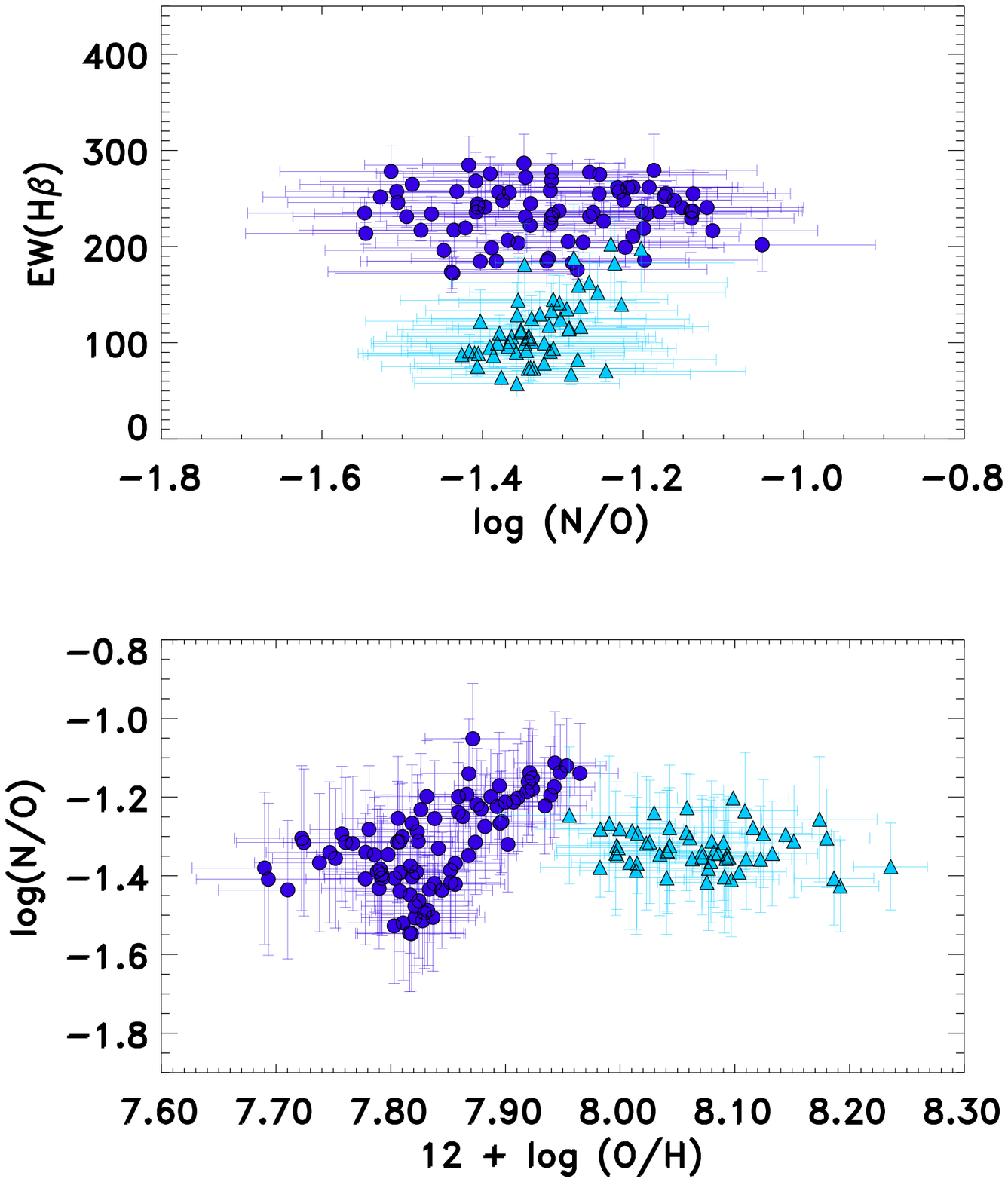} 
\caption{Top panel: log (N/O) ratio versus EW(H$\beta$). Bottom panel: 12 + log (O/H) ratio versus log(N/O).
Triangles correspond to the data points of Tol 0104-388 and circles corresponds
with the data points of Tol 2146-391. More details in \cite{Lagos12}.}
   \label{fig5}
\end{center}
\end{figure}

In summary, the results obtained in our studies suggest that the chemical properties (O, N and N/O) across 
H{\sc ii}/BCD galaxies are fairly uniform, although a slight gradient of O and N are observed in the 
ISM of UM 408 and Tol 2146-391, respectively. 
We suggest that global hydro-dynamical processes, such as starburst-driven super-shells or/and inflow of gas 
might be governing the transport and mixing of metals across these galaxies,
keeping the N/O ratio constant through the ISM at large scales \cite{PerezMontero11,Lagos12}.

\section{The high ionization emission line He \,{\sc ii} $\lambda$4686}\label{hard}

The origin of high-ionization emission lines, such as [Ne V] $\lambda$3426, 
[Fe V] $\lambda$4227 and He \,{\sc ii} $\lambda$4686 in starburst and H\,{\sc ii}/BCD galaxies has been 
a subject of study in the last years, given that photoionization models of H\,{\sc ii} regions generally 
fail to reproduce the observed intensities of these lines (see, e.g., \cite{Izotov04}).
Several mechanisms for producing hard ionizing radiation have been proposed in the literature, such as 
WR stars \cite{Schaerer96}, primordial (zero-metallicity) stars, high-mass X-ray binaries (HMXB; \cite{Garnett91}),  
radiative shocks \cite{DopitaSutherland96} and O stars at low metallicity \cite{Brinchmann08}.
In \cite{Lagos12} and \cite{Lagos13} we studied with GMOS-IFU the spatial distribution of 
He \,{\sc ii} $\lambda$4686 in the compact H\,{\sc ii} galaxies Tol\ 0104-388 and Tol\ 2146-391, 
and in the XMP BCD galaxy HS\ 2236+1344, respectively, in order to gain insights into the nature 
of their hard ionization radiation and its possible dependence on the properties of the ISM  
\cite{Garnett91,Izotov04}.

Based on a spaxel-by-spaxel analysis, instead of the integrated properties of the galaxies 
(see Figure 15 in \cite{Lagos12}),
our results indicate that the spatial distribution of He \,{\sc ii} $\lambda$4686 relative to H$\beta$ does not 
depend on the EW(H$\beta$), oxygen abundance or log(N/O).
In particular, the oxygen abundance appears to be constant through 
the whole extent of our sample galaxies, as already is observed in other H\,{\sc ii}/BCD galaxies
(e.g., \cite{Lagos09}; and references therein).
The opposite trend is found if we consider the integrated spectra of galaxies, in the sense that 
this emission line is stronger in galaxies at low metallicity \cite{Izotov06b}.
The lack of a relationship between the hardness of the ionizing radiation and the EW(H$\beta$), or age, 
\cite{ThuanIzotov05} suggests that the presence of high-ionization lines, in particular 
He \,{\sc ii} $\lambda$4686, is not due to a single excitation mechanism.
For instance, in \cite{Guseva00}, it was found for a sample of galaxies, 
with detected and non-detected WR features, the same dependence of 
I(He \,{\sc ii} $\lambda$4686)/I(H$\beta$) on the EW(H$\beta$). 
This indicates that WR stars are not the sole origin of He \,{\sc ii} $\lambda$4686 in star-forming regions
(see also \cite{Shirazi12}). 
In galaxies with detected WR stars, the He \,{\sc ii} $\lambda$4686 commonly appears to not be coincident 
with the location of the WR bumps (e.g., Mrk 178)
and these stellar features are not always seen when nebular He \,{\sc ii} is observed (e.g., Tol 2146-391, 
HS2236+1344, Tol 65). The spatial offset between WR stars and He \,{\sc ii} $\lambda$4686, in Mrk 178, 
is interpreted by \cite{Kehrig13} as an effect of the mechanical energy injected by WR star winds, so 
WR stars are not ruled out as the main source of the observed He \,{\sc ii} $\lambda$4686 in that galaxy. 
An examination of individual spaxels in our data cubes, and also in the integrated spectra
of our sample galaxies (e.g., Figure \ref{fig1} in this contribution), does not reveal any clear stellar WR features.
In the case of the XMP BCD galaxy HS\ 2236+1344, we detected the He\,{\sc ii} $\lambda$4686 emission line 
in only one of the GH\,{\sc ii}Rs of the galaxy (the brightest one).
In this galaxy, the He \,{\sc ii} $\lambda$4686 line appears to be excited through point sources within 
a compact volume which, interestingly, does not coincide with the position where a high N/O abundance ratio 
has been observed. We discuss, in \cite{Lagos13}, the possibility that the He\,{\sc ii} $\lambda$4686 emission line, 
in HS2236+1344, is associated with WR stars, high-mass X-ray binaries (HMXBs), O stars at low metallicities, 
and/or a low-luminosity Active Galactic Nucleus. However, since clear WR features have not been 
detected in that galaxy, WR stars are excluded as the primary excitation source of 
He \,{\sc ii} $\lambda$4686 emission.

\section{Conclusions}\label{conclusion}

As far as the spatial distribution of oxygen abundances is concerned,  
we did not detect localized overabundances in any of our sample galaxies. 
However, we find evidence for a marginal negative radial abundance gradient, with the 
highest abundances seen at the position of the brightest star cluster complexes 
(peak of H$\alpha$ emission), in the H\,{\sc ii}/BCD galaxy UM\ 408 at least.
If real, the slight trend for an increasing 12+log(N/H) abundance, in the galaxy Tol 2146-391, suggests
rapid self-enrichment by the freshly produced heavy elements in the present starburst on scales of hundreds of pc, or, 
alternatively, metal pollution by a previous star formation episode.
In any case, the oxygen and nitrogen appear to be well mixed across the ISM of H\,{\sc ii}/BCD galaxies, suggesting 
efficient transport by expanding starburst-driven supershells and/or gas infall from the halo.

Our spectroscopic IFU studies suggest a mixture of compact sources as the main excitation source for localized 
He \,{\sc ii} $\lambda$4686 emission in H\,{\sc ii}/BCD galaxies, without clear WR signatures,
with WR stars probably being of secondary importance. In the galaxy Tol 2146-391, we favor the idea of extended 
He \,{\sc ii} $\lambda$4686 emission being primarily due to radiative shocks in the ISM.

\section{Acknowledgments}

P.L. is supported by a Post-Doctoral grant SFRH/BPD/72308/2010, 
funded by FCT (Portugal), and P.P. by Ciencia 2008 Contract, funded by FCT/MCTES 
(Portugal) and POPH/FSE (EC).  
We are very thankful to Andrew Humphrey for their very useful
suggestions which have improved the paper.
We would like thank the anonymous referee for his/her comments and suggestions which substantially improved the paper.
We acknowledge support by the Funda\c{c}\~{a}o para a Ci\^{e}ncia e a Tecnologia (FCT)
under project FCOMP-01-0124-FEDER-029170 (Reference FCT PTDC/FIS-AST/3214/2012),
funded by the FEDER program. 
This research has made use of the NASA/IPAC Extragalactic
Database (NED) which is operated by the Jet Propulsion laboratory, California
Institute of technology, under contract with the National Aeronautics and Space
Administration. We acknowledge the usage of the HyperLeda database (http://leda.univ-lyon1.fr).
The data presented in this paper have been obtained through the  
Gemini programs GS-2004B-Q-59, GS-2005B-Q-19 and GN-2010B-Q-69, and the ESO-VLT program 090.B-0242. 

%

\end{document}